\documentclass[12pt]{iopart}

\usepackage{cite}
\usepackage{graphicx}

\begin{document}

\title[]{Accurate Modelling of Left-Handed Metamaterials Using Finite-Difference
Time-Domain Method with Spatial Averaging at the Boundaries}

\author{Yan Zhao, Pavel Belov and Yang Hao}

\address{Electronic Engineering, Queen Mary, University of London, London E1 4NS, UK}
\ead{yan.zhao@elec.qmul.ac.uk}

\begin{abstract}
The accuracy of finite-difference time-domain (FDTD) modelling of
left-handed metamaterials (LHMs) is dramatically improved by using
an averaging technique along the boundaries of LHM slabs. The
material frequency dispersion of LHMs is taken into account using
auxiliary differential equation (ADE) based dispersive FDTD methods.
The dispersive FDTD method with averaged permittivity along the
material boundaries is implemented for a two-dimensional (2-D)
transverse electric (TE) case. A mismatch between analytical and
numerical material parameters (e.g. permittivity and permeability)
introduced by the time discretisation in FDTD is demonstrated. The
expression of numerical permittivity is formulated and it is
suggested to use corrected permittivity in FDTD simulations in order
to model LHM slabs with their desired parameters. The influence of
switching time of source on the oscillation of field intensity is
analysed. It is shown that there exists an optimum value which leads
to fast convergence in simulations.
\end{abstract}

\maketitle

\section{Introduction}
Recently a great attention has been paid on the research of a new
type of artificial materials: medium with simultaneously negative
permittivity and permeability which is introduced by Veselago in his
early paper in 1968 \cite{Veselago} and named as left-handed
metamaterial (LHM). The electric field, magnetic field and wave
vector of an electromagnetic plane wave in such materials form a
left-handed system of vectors. The LHMs introduce peculiar yet
interesting properties such as negative refraction, reversed Doppler
effect and reversed Cerenkov radiation etc. One of the most
important applications of LHMs suggested by Sir John Pendry, is the
``perfect lens'' \cite{Pendrylens}, e.g. the subwavelength imaging
which allows the information below the diffraction limit of
conventional imaging systems to be transported. The LHM lenses
provide unique properties of negative refraction and amplification
of evanescent waves, which accounts for the reconstruction of source
information at the image plane.

The finite-difference time-domain (FDTD) method \cite{Taflove} is a
versatile and robust technique. Over the years it has been widely
used for the modelling of electromagnetic wave interaction with
various frequency dispersive and non-dispersive materials. For
modelling of LHMs with negative material properties, the frequency
dispersion has to be taken into account therefore the dispersive
FDTD method needs to be used. The existing frequency dispersive FDTD
methods can be categorised into three types: the recursive
convolution (RC) method \cite{Luebbers1}, the auxiliary differential
equation (ADE) method \cite{Gandhi1} and the $Z$-transform method
\cite{Sullivan1}. The RC scheme relates electric flux density to
electric field intensity through a convolution integral, which can
be discretised as a running sum. The dispersive FDTD method applying
the RC scheme has been used for modelling of different types of
dispersive materials in
\cite{Luebbers2,Luebbers3,Hunsberger,Melon,Akyurtlu1,Grande,Akyurtlu2,Lee}.
The ADE method introduces additional differential equations in order
to describe frequency dependent material properties
\cite{Kashiwa1,Kashiwa2,Goorjian,Gandhi2,Gandhi3,Lu}. Another
dispersive FDTD method is based on the $Z$-transforms
\cite{Sullivan2,Demir}: the time-domain convolution integral is
reduced to a multiplication using the $Z$-transform, and a recursive
relation between electric flux density and electric field is
derived.

There have been a number of attempts to model LHMs using the FDTD
method \cite{ZiolkowskiPRE,Loschialpo,Cummer,Rao,Feise,LimitFDTD}.
It may seem that the conventional dispersive FDTD has been verified
in the literature: the negative refraction effect which is inherent
to the boundary between the free space and LHM is observed and the
planar superlens behaviour has been successfully demonstrated
\cite{ZiolkowskiPRE,Loschialpo,Cummer}. Actually, this means that
the LHM is correctly modelled only for the case of propagating
waves. When evanescent waves are considered the conventional
implementation of dispersive FDTD method may lead to inaccurate
results. Usually, the evanescent waves decay exponentially over
distances and thus they are concentrated in the close vicinity of
sources, that is why conventional FDTD modelling of non-dispersive
materials does not suffer from the aforementioned numerical
inaccuracy. In the case of LHM, the evanescent waves play a key role
and have to be modelled accurately because of the perfect lens
effect \cite{Pendrylens}. This explains why early FDTD simulations
have not demonstrated the subwavelength imaging property of LHM
lenses \cite{ZiolkowskiPRE,Loschialpo}. A slab of LHM effectively
amplifies evanescent waves which normally decay in usual materials
and allows transmission of subwavelength details of sources to
significant distances.

Other numerical studies using the FDTD method include the effect of
losses and thicknesses on the transmission characteristics of LHM
slabs \cite{Rao}, and the influence of numerical material parameters
on their imaging properties \cite{LimitFDTD} etc. Besides the FDTD
method, the pseudo-spectral time-domain (PSTD) method has been used
for the modelling of backward-wave metamaterials
\cite{FDTDLHMFeise}. It is claimed in \cite{FDTDLHMFeise} that the
FDTD method cannot be used to accurately model LHMs due to the
numerical artefact of the staggered grid in FDTD domain. However, we
shall show later by comparing the transmission coefficient
calculated from FDTD simulation and exact analytical solutions that
with proper field averaging techniques \cite{LimitFDTD,YanPRL}, the
FDTD method indeed can be used to accurately characterise the
behavior of both propagating and evanescent waves in LHM slabs.
Furthermore, it has been reported in
\cite{MohammadiCurve,MohammadiFlat} that with special treatment
(i.e. averaging techniques) along material boundaries, accurate
modelling of curved surfaces of conventional dielectrics as well as
surface plasmon polaritons between metal-dielectric interfaces can
be achieved without using extremely fine FDTD meshes.

Ideally lossless LHM slabs with infinite transverse length provide
unlimited subwavelength resolution. However in realistic situations,
the subwavelength resolution of the LHM lenses is limited by losses
\cite{Podolskiy}, the thickness of the slab and the mismatch of the
slab with its surrounding medium \cite{SmithLimit}. It is important
to understand these theoretical limitations because they can help
verify numerical simulations. In this paper, we have performed the
modelling of infinite LHM slabs and their transmission
characteristics. The infinite LHM slab is modelled using the
periodic boundary condition and a material parameter averaging
technique is used along the boundaries of LHM slabs. In contrast to
FDTD modelling of conventional dielectric slabs where the averaging
is only a second-order correction to improve the accuracy of
simulations, the averaging of permittivity is an essential
modification for modelling of LHM slabs. The averaging of material
parameters implemented in our FDTD simulations is equivalent to the
averaging of current density originally introduced in
\cite{LimitFDTD} and is analysed in detail in this paper. It is
demonstrated that other numerical aspects such as numerical material
parameters and the switching time of source also have considerable
influences on FDTD simulations.

\section{Dispersive FDTD Modelling of LHMs with Spatial Averaging at the Boundaries}
We consider here lossy isotropic LHM slabs modelled using the
effective medium method. The Drude model is used for both the
permittivity $\varepsilon(\omega)$ and permeability $\mu(\omega)$
with identical dispersion forms:
\begin{eqnarray}
\varepsilon(\omega)=\varepsilon_0\left(1-\frac{\omega_{pe}^{2}}{\omega^{2}-j\omega\gamma_e}\right),
\label{eq_permittivity}
\end{eqnarray}
\begin{eqnarray}
\mu(\omega)=\mu_0\left(1-\frac{\omega_{pm}^{2}}{\omega^{2}-j\omega\gamma_m}\right),
\label{eq_permittivity_m}
\end{eqnarray}
where $\omega_{pe}$ and $\omega_{pm}$ are electric and magnetic
plasma frequencies and $\gamma_e$ and $\gamma_m$ are electric and
magnetic collision frequencies, respectively.

Although there are various dispersive FDTD methods available for the
modelling of LHMs, due to its simplicity and efficiency, we have
implemented the ADE method in this paper. There are also different
schemes involving different auxiliary differential equations in
addition to conventional FDTD updating equations. In this paper, two
schemes, namely the (\textbf{E}, \textbf{J}, \textbf{H}, \textbf{M})
scheme \cite{Taflove} and the (\textbf{E}, \textbf{D}, \textbf{H},
\textbf{B}) scheme \cite{Gandhi1}, are used and introduced
respectively.

\subsection{The (\textbf{E}, \textbf{D}, \textbf{H}, \textbf{B}) Scheme}
The (\textbf{E}, \textbf{D}, \textbf{H}, \textbf{B}) scheme is based
on Faraday's and Ampere's Laws:
\begin{eqnarray}
\textrm{curl}(\textbf{E})&=&-\frac{\partial\textbf{B}}{\partial t},
\label{eq_Maxwell_E}\\
\textrm{curl}(\textbf{H})&=&\frac{\partial\textbf{D}}{\partial t},
\label{eq_Maxwell_H}
\end{eqnarray}
as well as the constitutive relations
$\textbf{D}=\varepsilon\textbf{E}$ and $\textbf{B}=\mu\textbf{H}$
where $\varepsilon$ and $\mu$ are expressed by
(\ref{eq_permittivity}) and (\ref{eq_permittivity_m}), respectively.
Equations (\ref{eq_Maxwell_E}) and (\ref{eq_Maxwell_H}) can be
discretised following a normal procedure \cite{Taflove} which leads
to conventional FDTD updating equations:
\begin{eqnarray}
\textbf{B}^{n+1}&=&\textbf{B}^n-\Delta
t\cdot\overline{\textrm{curl}}(\textbf{E}^{n+\frac{1}{2}}),
\label{eq_Maxwell_B_approx}\\
\textbf{D}^{n+1}&=&\textbf{D}^n+\Delta
t\cdot\overline{\textrm{curl}}(\textbf{H}^{n+\frac{1}{2}}).
\label{eq_Maxwell_D_approx}
\end{eqnarray}
where $\overline{\textrm{curl}}$ is discrete curl operator, $\Delta
t$ is FDTD time step and $n$ is the number of time steps.

In addition, auxiliary differential equations have to be taken into
account and they can be discretised through the following steps. The
constitutive relation between \textbf{D} and \textbf{E} reads
\begin{equation}
\left(\omega^2-j\omega\gamma_e\right)\textbf{D}=\varepsilon_0\left(\omega^2-j\omega\gamma_e-\omega^2_{pe}\right)\textbf{E}.
\label{eq_constitutive}
\end{equation}
Using inverse Fourier transform and the following rules:
\begin{equation}
j\omega\rightarrow\frac{\partial}{\partial
t},~~~~\omega^2\rightarrow-\frac{\partial^2}{\partial t^2},
\label{eq_inverse_Fourier}
\end{equation}
Equation (\ref{eq_constitutive}) can be rewritten in the time domain
as
\begin{equation}
\left(\frac{\partial^2}{\partial t^2}+\frac{\partial}{\partial
t}\gamma_e\right)\textbf{D}=\varepsilon_0\left(\frac{\partial^2}{\partial
t^2}+\frac{\partial}{\partial
t}\gamma_e+\omega^2_{pe}\right)\textbf{E}. \label{eq_constitutive_t}
\end{equation}

The FDTD simulation domain is represented by an equally spaced
three-dimensional (3-D) grid with periods $\Delta x$, $\Delta y$ and
$\Delta z$ along $x$-, $y$- and $z$-directions, respectively. For
discretisation of (\ref{eq_constitutive_t}), we use central finite
difference operators in time ($\delta_t$ and $\delta^2_t$) and
central average operator with respect to time ($\mu_t$ and
$\mu^2_t$):
\begin{eqnarray}
\frac{\partial^2}{\partial t^2}\rightarrow\frac{\delta^2_t}{(\Delta
t)^2},~~~~\frac{\partial}{\partial
t}\rightarrow\frac{\delta_t}{\Delta
t}\mu_t,~~~~\omega^2_{pe}\rightarrow\omega^2_{pe}\mu^2_t,\nonumber
\label{eq_operator}
\end{eqnarray}
where the operators $\delta_t$, $\delta^2_t$, $\mu_t$ and $\mu^2_t$
are defined as in \cite{Hildebrand}:
\begin{eqnarray}
\delta_t\textbf{F}|^n_{m_x,m_y,m_z}&\equiv&\textbf{F}|^{n+\frac{1}{2}}_{m_x,m_y,m_z}-\textbf{F}|^{n-\frac{1}{2}}_{m_x,m_y,m_z}\\
\delta^2_t\textbf{F}|^n_{m_x,m_y,m_z}&\equiv&\textbf{F}|^{n+1}_{m_x,m_y,m_z}-2\textbf{F}|^n_{m_x,m_y,m_z}+\textbf{F}|^{n-1}_{m_x,m_y,m_z}\nonumber\\
\mu_t\textbf{F}|^n_{m_x,m_y,m_z}&\equiv&\frac{\textbf{F}|^{n+\frac{1}{2}}_{m_x,m_y,m_z}+\textbf{F}|^{n-\frac{1}{2}}_{m_x,m_y,m_z}}{2}\nonumber\\
\mu^2_t\textbf{F}|^n_{m_x,m_y,m_z}&\equiv&\frac{\textbf{F}|^{n+1}_{m_x,m_y,m_z}+2\textbf{F}|^n_{m_x,m_y,m_z}+\textbf{F}|^{n-1}_{m_x,m_y,m_z}}{4}\nonumber
\label{eq_operators}
\end{eqnarray}
Here $\textbf{F}$ represents field components and
$m_{x},m_{y},m_{z}$ are indices corresponding to a certain
discretisation point in FDTD domain. The discretised Eq.
(\ref{eq_constitutive_t}) reads
\begin{equation}
\left[\frac{\delta^2_t}{(\Delta t)^2}+\frac{\delta_t}{\Delta
t}\mu_t\gamma_e\right]\textbf{D}=\varepsilon_0\left[\frac{\delta^2_t}{(\Delta
t)^2}+\frac{\delta_t}{\Delta
t}\mu_t\gamma_e+\omega^2_{pe}\mu^2_t\right]\textbf{E}.
\label{eq_constitutive_approx}
\end{equation}
Note that in (\ref{eq_constitutive_approx}), the discretisation of
term $\omega^2_{pe}$ of (\ref{eq_constitutive_t}) is performed using
the central average operator $\mu^2_t$ in order to guarantee
improved stability; the central average operator $\mu_t$ is used for
the term containing $\gamma_e$ to preserve second-order feature of
the equation. Equation (\ref{eq_constitutive_approx}) can be written
as
\begin{eqnarray}
\lefteqn{\frac{\textbf{D}|^{n+1}_{m_x,m_y,m_z}-2\textbf{D}|^n_{m_x,m_y,m_z}+\textbf{D}|^{n-1}_{m_x,m_y,m_z}}{(\Delta
t)^2}+\gamma_e\frac{\textbf{D}|^{n+1}_{m_x,m_y,m_z}-\textbf{D}|^{n-1}_{m_x,m_y,m_z}}{2\Delta
t}}\nonumber\\
&&\!\!\!\!\!\!\!=\varepsilon_0\Biggl[\frac{\textbf{E}|^{n+1}_{m_x,m_y,m_z}-2\textbf{E}|^n_{m_x,m_y,m_z}+\textbf{E}|^{n-1}_{m_x,m_y,m_z}}{(\Delta
t)^2}+\gamma_e\frac{\textbf{E}|^{n+1}_{m_x,m_y,m_z}-\textbf{E}|^{n-1}_{m_x,m_y,m_z}}{2\Delta
t}\nonumber\\
&&~~~~+\omega^2_{pe}\frac{\textbf{E}|^{n+1}_{m_x,m_y,m_z}+2\textbf{E}|^n_{m_x,m_y,m_z}+\textbf{E}|^{n-1}_{m_x,m_y,m_z}}{4}\Biggr].
\label{eq_constitutive_approx2}
\end{eqnarray}
Therefore the updating equation for \textbf{E} in terms of
\textbf{E} and \textbf{D} at previous time steps is as follows:
\begin{eqnarray}
\lefteqn{\textbf{E}^{n+1}=\Bigg\{\left[\frac{1}{\varepsilon_0(\Delta
t)^2}+\frac{\gamma_e}{2\varepsilon_0\Delta
t}\right]\textbf{D}^{n+1}-\frac{2}{\varepsilon_0(\Delta
t)^2}\textbf{D}^n}\nonumber\\
&&~~~~~~~~~~+\left[\frac{2}{(\Delta
t)^2}-\frac{\omega^2_{pe}}{2}\right]\textbf{E}^n
-\left[\frac{1}{(\Delta t)^2}-\frac{\gamma_e}{2\Delta
t}+\frac{\omega^2_{pe}}{4}\right]\textbf{E}^{n-1}
\nonumber\\
&&~~~~~~~~~~+\left[\frac{1}{\varepsilon_0(\Delta
t)^2}-\frac{\gamma_e}{2\varepsilon_0\Delta t}\right]\textbf{D}^{n-1}
\Bigg\}\Bigg/\left[\frac{1}{(\Delta t)^2}+\frac{\gamma_e}{2\Delta
t}+\frac{\omega^2_{pe}}{4}\right], \label{eq_DE}
\end{eqnarray}

The updating equation for \textbf{H} is in the same form as
(\ref{eq_DE}) by replacing \textbf{E}, \textbf{D}, $\omega^2_{pe}$
and $\gamma_e$ by \textbf{H}, \textbf{B}, $\omega^2_{pm}$ and
$\gamma_m$, respectively i.e.
\begin{eqnarray}
\lefteqn{\textbf{H}^{n+1}=\Bigg\{\left[\frac{1}{\varepsilon_0(\Delta
t)^2}+\frac{\gamma_m}{2\varepsilon_0\Delta
t}\right]\textbf{B}^{n+1}-\frac{2}{\varepsilon_0(\Delta
t)^2}\textbf{B}^n}\nonumber\\
&&~~~~~~~~~~+\left[\frac{2}{(\Delta
t)^2}-\frac{\omega^2_{pm}}{2}\right]\textbf{H}^n-\left[\frac{1}{(\Delta
t)^2}-\frac{\gamma_m}{2\Delta
t}+\frac{\omega^2_{pm}}{4}\right]\textbf{H}^{n-1}\nonumber\\
&&~~~~~~~~~~+\left[\frac{1}{\varepsilon_0(\Delta
t)^2}-\frac{\gamma_m}{2\varepsilon_0\Delta
t}\right]\textbf{B}^{n-1}\Bigg\}\Bigg/\left[\frac{1}{(\Delta
t)^2}+\frac{\gamma_m}{2\Delta t}+\frac{\omega^2_{pm}}{4}\right].
\label{eq_BH}
\end{eqnarray}
Equations (\ref{eq_Maxwell_B_approx}), (\ref{eq_Maxwell_D_approx}),
(\ref{eq_DE}) and (\ref{eq_BH}) form an FDTD updating equation set
for LHMs using the (\textbf{E}, \textbf{D}, \textbf{H}, \textbf{B})
scheme. If both the plasma frequency and collision frequency are
equal to zero i.e. $\omega_{pe}=\omega_{pm}=0$ and
$\gamma_e=\gamma_m=0$, then they reduce to the updating equations in
the free space.

\subsection{The (\textbf{E}, \textbf{J}, \textbf{H}, \textbf{M}) Scheme}
An alternative ADE FDTD scheme starts with different forms of
Faraday's and Ampere's Laws for LHMs:
\begin{eqnarray}
\textrm{curl}(\textbf{E})&=&-\mu_0\frac{\partial\textbf{H}}{\partial
t}-\textbf{M},
\label{eq_Maxwell_E1}\\
\textrm{curl}(\textbf{H})&=&\varepsilon_0\frac{\partial\textbf{E}}{\partial
t}+\textbf{J}, \label{eq_Maxwell_H1}
\end{eqnarray}
where the electric and magnetic current density, \textbf{J} and
\textbf{M} are defined as
\begin{eqnarray}
\textbf{J}(\omega)&=&j\omega\varepsilon_0\frac{\omega^2_{pe}}{j\omega\gamma_e-\omega^2}\textbf{E}(\omega),
\label{eq_J}\\
\textbf{M}(\omega)&=&j\omega\varepsilon_0\frac{\omega^2_{pm}}{j\omega\gamma_m-\omega^2}\textbf{H}(\omega).
\label{eq_M}
\end{eqnarray}
Following the same procedure as for the (\textbf{E}, \textbf{D},
\textbf{H}, \textbf{B}) scheme, Eqs.
(\ref{eq_Maxwell_E1})-(\ref{eq_M}) can be discretised as:
\begin{eqnarray}
\textbf{H}^{n+1}&=&\textbf{H}^n-\frac{\Delta
t}{\mu_0}\left[\overline{\textrm{curl}}(\textbf{E}^{n+\frac{1}{2}})+\textbf{M}^{n+\frac{1}{2}}\right],
\label{eq_Maxwell_H1_approx}\\
\textbf{E}^{n+1}&=&\textbf{E}^n+\frac{\Delta
t}{\varepsilon_0}\left[\overline{\textrm{curl}}(\textbf{H}^{n+\frac{1}{2}})-\textbf{J}^{n+\frac{1}{2}}\right],
\label{eq_Maxwell_E1_approx}
\end{eqnarray}
\begin{eqnarray}
\lefteqn{\textbf{J}|^{n+1}_{m_x,m_y}=\frac{4}{\gamma_e\Delta
t+2}\textbf{J}|^n_{m_x,m_y}+\frac{\gamma_e\Delta t-2}{\gamma_e\Delta
t+2}\textbf{J}|^{n-1}_{m_x,m_y}}\nonumber\\
&&~~~~~~~~~~~~~+\frac{\varepsilon_0\omega^2_{pe}\Delta
t}{\gamma_e\Delta
t+2}\left(\textbf{E}|^{n+1}_{m_x,m_y}-\textbf{E}|^{n-1}_{m_x,m_y}\right),
\label{eq_J_approx}\\
\lefteqn{\textbf{M}|^{n+1}_{m_x,m_y}=\frac{4}{\gamma_m\Delta
t+2}\textbf{M}|^n_{m_x,m_y}+\frac{\gamma_m\Delta t-2}{\gamma_m\Delta
t+2}\textbf{M}|^{n-1}_{m_x,m_y}}\nonumber\\
&&~~~~~~~~~~~~~~+\frac{\varepsilon_0\omega^2_{pm}\Delta
t}{\gamma_m\Delta
t+2}\left(\textbf{H}|^{n+1}_{m_x,m_y}-\textbf{H}|^{n-1}_{m_x,m_y}\right).
\label{eq_M_approx}
\end{eqnarray}
Again Eqs. (\ref{eq_Maxwell_H1_approx})-(\ref{eq_M_approx}) become
the free space updating equations if both the plasma frequency and
collision frequency are equal to zero i.e.
$\omega_{pe}=\omega_{pm}=0$ and $\gamma_e=\gamma_m=0$.

\subsection{The Spatial Averaging Methods}
\begin{figure}[t]
\centering
\includegraphics[width=7.6cm]{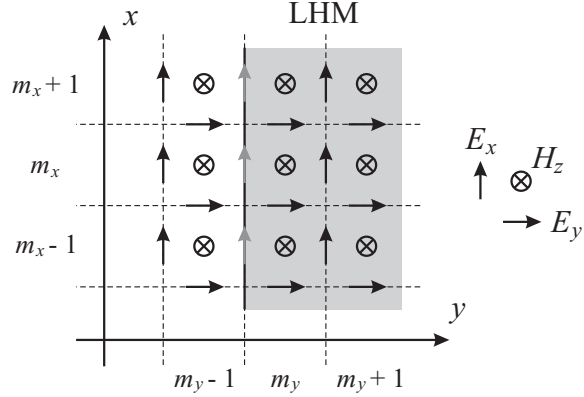}
\caption{The layout of FDTD grid illustrating the arrangement of
material boundaries along $y$-direction. The gray arrows indicate
where the averaged permittivity is used. The FDTD unit cell is shown
on the right side.} \label{fig_grid}
\end{figure}

In addition to the above introduced ADE schemes, due to the
staggered grid in FDTD domain, a modification at the interfaces
between different materials is often used to improve the accuracy of
FDTD simulations. It has been shown that the field averaging
techniques based on the averaging of material parameters (e.g.
permittivity and permeability) provide a second-order accuracy
\cite{Hwang}. The averaged permittivity/permeability can be obtained
by performing either arithmetic mean, harmonic mean or geometrical
mean \cite{Hwang} and the arithmetic mean has been proven to have
the best performance amount these three schemes. Previous analysis
of averaging techniques are performed for conventional dielectrics
with positive permittivity and permeability. For the materials with
negative permittivity/permeability, one of the simplest ways to
implement averaging is to use arithmetic mean. Furthermore,
averaging should be applied only for the field components tangential
to material interfaces. Therefore depending on the configuration of
FDTD simulation domain e.g. two-dimensional (2-D) TE, 2-D TM or
three-dimensional (3-D) cases, the averaging needs to be performed
in different ways. In this paper we have considered a 2-D ($x$-$y$)
simulation domain with $\textbf{H}$-polarisation where $\textbf{H}$
is directed only along $z$-direction. Therefore only three field
components are non-zero: $E_x$, $E_y$ and $H_z$. For the interfaces
between LHM slab and the free space along $x$-direction, the
averaged permittivity for the tangential electric field component
$E_x$ is given by
\begin{equation}
<\varepsilon_x>=\frac{\varepsilon_0+\varepsilon_x}{2}=\varepsilon_0\left[1-\frac{\omega^{2}_{pe}}{2\left(\omega^2-j\omega\gamma_e\right)}\right],
\label{eq_averaging}
\end{equation}
which is equivalent to replacing the plasma frequency $\omega_{pe}$
by $\omega'_{pe}=\omega_{pe}/\sqrt{2}$ in (\ref{eq_permittivity}).
Therefore along the boundaries, the updating equation for $E_x$
reads
\begin{eqnarray}
\lefteqn{E^{n+1}_x=\Bigg\{\left[\frac{1}{\varepsilon_0(\Delta
t)^2}+\frac{\gamma_e}{2\varepsilon_0\Delta
t}\right]D^{n+1}_x-\frac{2}{\varepsilon_0(\Delta
t)^2}D^n_x}\nonumber\\
&&~~~~~~~~~~+\left[\frac{2}{(\Delta
t)^2}-\frac{\omega^2_{pe}}{4}\right]E^n_x-\left[\frac{1}{(\Delta
t)^2}-\frac{\gamma_e}{2\Delta
t}+\frac{\omega^2_{pe}}{8}\right]E^{n-1}_x\nonumber\\
&&~~~~~~~~~~+\left[\frac{1}{\varepsilon_0(\Delta
t)^2}-\frac{\gamma_e}{2\varepsilon_0\Delta t}\right]D^{n-1}_x
\Bigg\}\Bigg/\left[\frac{1}{(\Delta t)^2}+\frac{\gamma_e}{2\Delta
t}+\frac{\omega^2_{pe}}{8}\right]. \label{eq_DEx2}
\end{eqnarray}
The locations where the updating equation (\ref{eq_DEx2}) is used is
illustrated in Fig. \ref{fig_grid} by gray arrows.

The averaging of permittivity can be implemented for the
(\textbf{E}, \textbf{D}, \textbf{H}, \textbf{B}) scheme. While for
the (\textbf{E}, \textbf{J}, \textbf{H}, \textbf{M}) scheme, it is
proposed in \cite{LimitFDTD} to use the averaging of tangential
current density along the boundaries of LHM slab. The averaged
current density can be calculated as (the free space current density
$J_0=0$):
\begin{equation}
<J_x>=\frac{J_0+J_x}{2}=\frac{J_x}{2}, \label{eq_averaging2}
\end{equation}
then the updating equation for $E_x$ along the boundaries of LHM
slab becomes (expanded from Eq. (\ref{eq_Maxwell_E1_approx}))
\begin{eqnarray}
\lefteqn{E_x|^{n+1}_{m_x,m_y}=\Bigg[E_x|^n_{m_x,m_y}+\frac{\Delta
t}{\varepsilon_0\Delta
y}\left(H_z|^{n+1}_{m_x,m_y}-H_z|^{n+1}_{m_x,m_y-1}\right)} \label{eq_ADE_E2}\\
&&~~~~~~~~~~~~~~~-\frac{\Delta t(\gamma_e\Delta
t+6)}{4\varepsilon_0(\gamma_e\Delta
t+2)}J_x|^n_{m_x,m_y}-\frac{\Delta t(\gamma_e\Delta
t-2)}{4\varepsilon_0(\gamma_e\Delta
t+2)}J_x|^{n-1}_{m_x,m_y} \nonumber\\
&&~~~~~~~~~~~~~~~+\frac{\omega^2_{pe}(\Delta t)^2}{4\gamma_e\Delta
t+8}E_x|^{n-1}_{m_x,m_y}\Bigg]\Bigg/\left(1+\frac{\omega^2_{pe}(\Delta
t)^2}{4\gamma_e\Delta t+8}\right).\nonumber
\end{eqnarray}

Theoretically the above two averaging methods have same effects due
to the linear relations
\begin{equation}
\textbf{D}=\varepsilon\textbf{E}=\varepsilon_0\textbf{E}+\frac{1}{j\omega}\textbf{J},~~~~\textbf{B}=\mu\textbf{H}=\mu_0\textbf{H}+\frac{1}{j\omega}\textbf{M}.
\label{eq_D}
\end{equation}
Therefore the averaging of current density is identical to the
averaging of permeability. In this paper, we have used the
(\textbf{E}, \textbf{D}, \textbf{H}, \textbf{B}) scheme in all our
simulations because of its simplicity in implementation. In order to
demonstrate the advantage of averaging technique, we have also
compared the results from simulations with and without averaged
permittivity along material boundaries. For the case of without
averaging, the tangential electric fields indicated by gray arrows
in Fig.~\ref{fig_grid} are updated using their updating equations in
the free space.

The above averaging of permittivity only applies to the field
components tangential to material interfaces and for the case of TE
polarisation considered in our simulations. If it is required to
apply the averaging schemes to materials with planar boundaries for
TM and three-dimensional (3-D) cases or even for structures with
curved surfaces, one can follow the procedures introduced in
\cite{MohammadiCurve,MohammadiFlat}.

\section{Numerical Implementation}
For simplicity, in our simulations we assume that the plasma
frequency is $\omega_{pe}=\omega_{pm}=\omega_p=\sqrt{2}\omega$ where
$\omega$ is the operating frequency, therefore matched LHM slabs are
modelled in our simulations. A small amount of losses is used i.e.
$\gamma_e=\gamma_m=\gamma=0.0005\omega$ which gives relative
permittivity and permeability $\varepsilon_r=\mu_r=-1-0.001j$ to
ensure the convergence of simulations. It is worth mentioning that
there is a small amount of mismatch between numerical (in FDTD
domain) and analytical permittivity (\ref{eq_permittivity}) which is
caused by FDTD time discretisation \cite{LimitFDTD}. However, such a
mismatch causes the amplification of transmission coefficient only
for lossless LHM slabs or when the losses are very small. For the
amount of losses used in our simulations, the effect of mismatch is
damped and no amplification is found in transmission coefficient.
The effect of FDTD cell size on this mismatch is analysed in later
sections.

\begin{figure}[t]
\centering
\includegraphics[width=9cm]{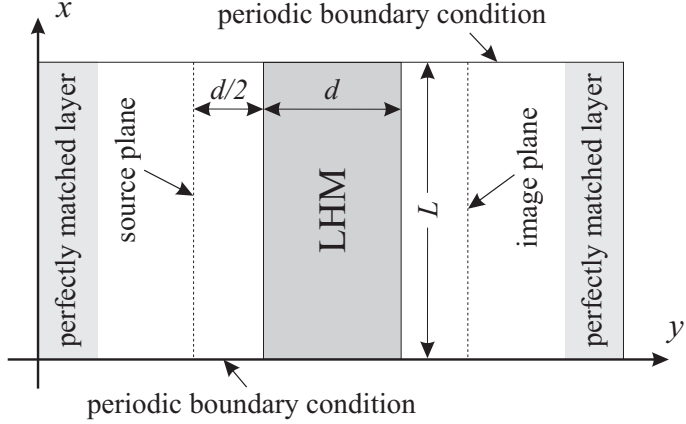}
\caption{Schematic diagram of two-dimensional (2-D) FDTD simulation
domain for calculation of numerical transmission coefficient}
\label{fig_domain_t}
\end{figure}

As shown in Fig. \ref{fig_domain_t}, an infinite LHM slab is
modelled by applying the Bloch's periodic boundary conditions
(PBCs).
For any periodic structures, the field at any time satisfies the
Bloch theory, i.e.
\begin{equation}
\textbf{E}(x+L)=\textbf{E}(x)e^{jk_xL},~~~~\textbf{H}(x+L)=\textbf{H}(x)e^{jk_xL},
\label{eq_PBC}
\end{equation}
where $x$ is any location in the computation domain, $k_x$ is the
wave number in $x$-direction and $L$ is the lattice period along the
direction of periodicity. When updating the fields at the boundary
of computation domain using the FDTD method, the required fields
outside the computation domain can be calculated using known field
values inside the domain through (\ref{eq_PBC}). Since infinite
structures can be truncated with any period, for saving computation
time, we have used only four FDTD cells in $x$-direction ($L=4\Delta
x$). Along $y$-direction, the Berenger's original perfectly matched
layer (PML) \cite{Berenger} is used for absorbing propagating waves
($k_x<k_0$), and the modified PML \cite{FangGPML} is used when
calculating the transmission coefficient for evanescent waves
($k_x>k_0$). A soft plane-wave sinusoidal source (which allows
scattered waves to pass through) with phase delay corresponding to
different wave number is used for excitations,
\begin{equation}
H_z\left(i,j_s\right)=H_z\left(i,j_s\right)+s(t)e^{-jk_x i\Delta x},
\label{eq_source}
\end{equation}
where $j_s$ is the location of source along $y$-direction, $s(t)$ is
a time domain sinusoidal wave function, $i\in\left[1,I\right]$ is
the index of cell location and $I$ is the total number of cells in
$x$-direction ($I=4$ in our case). By changing the values of wave
number $k_x$, either pure propagating waves ($k_x<k_0$) or pure
evanescent waves ($k_x>k_0$) can be excited.

The spatial resolution in FDTD simulations is $\Delta x=\Delta
y=\lambda/100$ where $\lambda$ is the free space wavelength at the
operating frequency. According to the stability criterion
\cite{Taflove}, the discretised time step is $\Delta t=\Delta
x/\sqrt{2}c$ where $c$ is the speed of light in the free space. As
illustrated in Fig. \ref{fig_domain_t}, the source plane is located
at a distance of $d/2$ to the front interface of the LHM slab where
$d$ is the thickness of the slab. Therefore the first image plane is
at the centre of the LHM slab and the second image plane is at the
same distance of $d/2$ beyond the slab. The spatial transmission
coefficient is calculated as a ratio of the field intensity at the
second image plane to the source plane for different transverse wave
numbers $k_x$ after the steady-state is reached in simulations.

\begin{figure}[t]
\centering
\includegraphics[width=10cm]{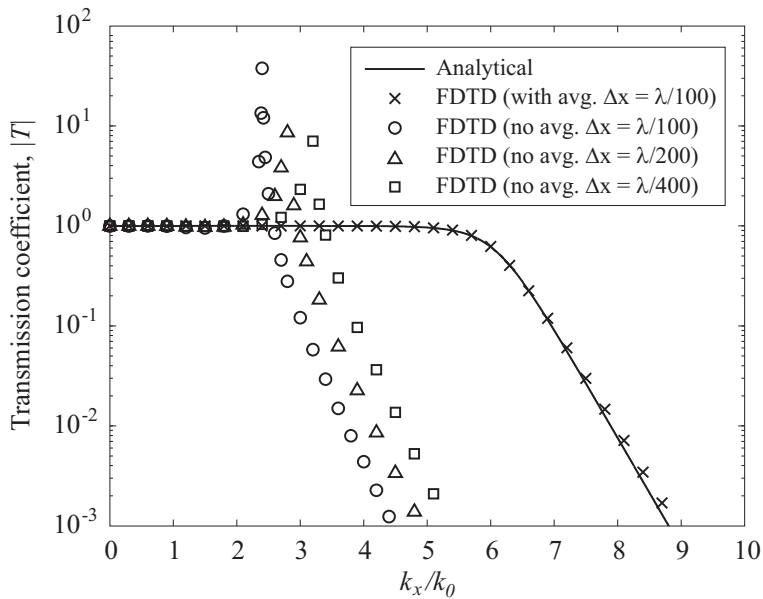}
\caption{Comparison of transmission coefficient of infinite planar
LHM slabs calculated from exact analytical solutions and dispersive
FDTD method with and without averaging of permittivity along the
boundaries of LHM slabs.} \label{fig_transmission1}
\end{figure}

Figure \ref{fig_transmission1} shows the transmission coefficient
for an infinite planar LHM slab with thickness $d=0.2\lambda$
calculated using the FDTD method with and without averaging of
permittivity along the boundaries, and its comparison with exact
analytical solutions.
It can be seen that using the arithmetic mean of permittivity, the
numerical results show excellent agreement with analytical solution
using spatial resolution $\Delta x=\lambda/100$. Good correspondence
can be also obtained when the FDTD cell size is increased to $\Delta
x=\lambda/80$. On the other hand, without averaging, the material
boundary is not correctly modelled which introduces an amplification
(resonance) at a location of approximately $k_x=2.4k_0$ in
transmission coefficient for the case of $\Delta x=\lambda/100$.
Reducing FDTD cell size to $\Delta x=\lambda/200$ and $\Delta
x=\lambda/400$, the behaviour of the resonance remains similar but
the location shifts to $k_x=2.8k_0$ and $k_x=3.2k_0$, respectively.
Therefore we predict that only if a very small FDTD cell size is
used in simulations that the results can converge to the right
solution. Such a comparison demonstrates the significance of the
averaging technique. Conventionally the arithmetic averaging is only
a second-order correction for modelling of conventional dielectric
slabs, however it is shown in Fig. \ref{fig_transmission1} that for
modelling of LHM slabs, the averaging becomes an essential
modification.

The results shown in Fig. \ref{fig_transmission1} may explain some
incorrect results obtained previously. For instance, the
amplification of transmission coefficient in \cite{Feise,Rao} is
caused by incorrect modelling of material boundaries, but such
amplification is pure numerical and does not exist in actual LHM
slabs \cite{SmithLimit,Podolskiy}. It is claimed in \cite{ChenPRL}
that the imaging property of finite-sized LHM slabs is significantly
affected by their transverse dimensions, which we suppose that the
conclusion is drawn from incorrect numerical simulations. We have
performed accurate simulations using averaging of material
properties and confirmed that the resolution of a near-field lens
using LHMs is free from its transverse aperture size \cite{YanPRL}.

In our simulations for the calculation of transmission coefficient,
we have used PBCs in $x$-direction to model infinite structures and
averaged permittivity along the boundaries in $y$-direction. If one
requires to model finite-sized structures (in both $x$- and
$y$-directions), the averaged permittivity/permeability needs to be
used for the corresponding tangential component along the boundaries
in both directions.

Besides the averaging technique used along the boundaries of LHM
slabs, there are other numerical aspects in FDTD simulations in
order to model the behaviour of LHM slab correctly and accurately.
These aspects are introduced respectively in following sections.

\section{Effects of Numerical Material Parameters}
Usually for modelling of conventional dielectrics, the results are
assumed to be accurate enough i.e. the effect of numerical material
parameters can be ignored if an FDTD cell size of smaller than
$\Delta x=\lambda/20$ is used. However since the discretisation
introduces a mismatch between numerical and analytical
permittivity/permeability, when modelling LHMs, especially when the
evanescent waves are involved, the FDTD spatial resolution has a
significant impact on the accuracy of simulation results. The effect
of numerical permittivity/permeability is originally reported in
\cite{LimitFDTD} for lossless LHMs using the (\textbf{E},
\textbf{J}, \textbf{H}, \textbf{M}) scheme. Following the same
procedure one can also obtain the numerical
permittivity/permeability for the case of lossy LHMs. In this paper,
the numerical permittivity/permeability (for lossy LHMs) is derived
for the (\textbf{E}, \textbf{D}, \textbf{H}, \textbf{B}) scheme.

In the case of plane waves, when
\begin{equation}
\textbf{E}^{n}=\textbf{E}e^{jn\omega\Delta
t},~~~~\textbf{D}^{n}=\textbf{D} e^{jn\omega\Delta t},
\label{eq_central difference}
\end{equation}
Eq. (\ref{eq_DE}) reduces to
$\textbf{D}^n=\tilde{\varepsilon}\textbf{E}^n$, where
$\tilde{\varepsilon}$ is the numerical permittivity of the following
form:
\begin{equation}
\tilde{\varepsilon}=\varepsilon_0\left[1-\frac{\omega^2_p(\Delta
t)^2\cos^2\frac{\omega\Delta t}{2}}{2\sin\frac{\omega\Delta
t}{2}\left(2\sin\frac{\omega\Delta t}{2}-j\gamma\Delta
t\cos\frac{\omega\Delta t}{2}\right)}\right]. \label{eq_epsilon}
\end{equation}
If the collision frequency $\gamma=0$, then (\ref{eq_epsilon})
reduces to the numerical permittivity for lossless LHMs given in
\cite{LimitFDTD}.

\begin{figure}[t]
\centering
\includegraphics[width=10cm]{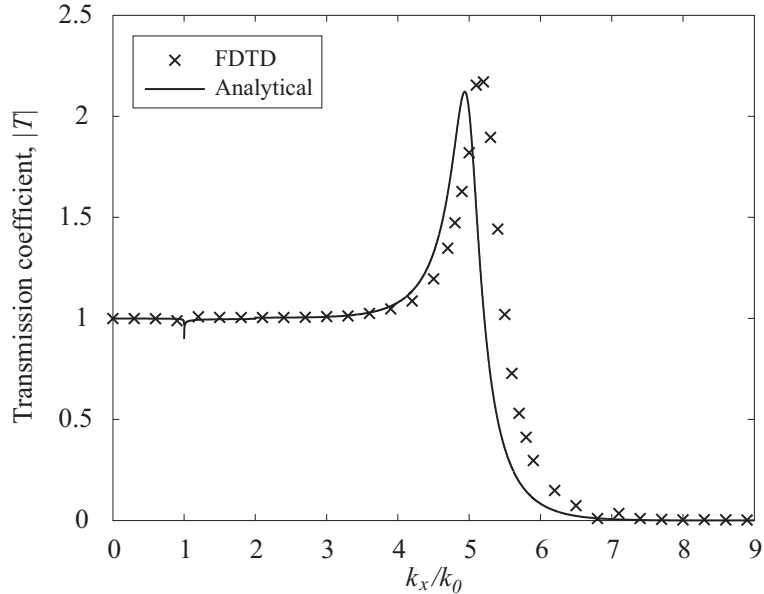}
\caption{Transmission coefficient of infinite planar LHM slabs using
the proposed FDTD method with averaged permittivity and without the
correction of material parameters. The amplification of transmission
coefficient is caused by the mismatch introduced by the time
discretisation in FDTD ($\varepsilon_r=-0.9959-0.0010j$) with
$\Delta x=\lambda/40$. The same permittivity is used to obtain the
analytical solution for comparison.} \label{fig_dispersion1}
\end{figure}

Previously we have used an FDTD cell size of $\Delta x=\lambda/100$
in simulations. Substitute the corresponding time step $\Delta
t=\Delta x/\sqrt{2}c$ and the operating frequency, we can obtain the
numerical relative permittivity from (\ref{eq_epsilon}) as
$\tilde{\varepsilon}_r=-0.9993-0.0010j$. Although there is a
mismatch between the real part of relative permittivity and $-1$,
the loss in LHMs damps such a mismatch and the simulation results
show very good accuracy. However if we increase FDTD cell size, the
numerical permittivity introduces severer mismatch which causes the
discrepancy between FDTD simulation result and exact solutions. For
example, for the case of $\Delta x=\lambda/40$, the mismatch brings
an amplification in transmission coefficient as shown in
Fig.~\ref{fig_dispersion1}.
Again using (\ref{eq_epsilon}) we can estimate this mismatch and the
numerical relative permittivity reads
$\tilde{\varepsilon}_r=-0.9959-0.0010j$. Using such permittivity in
analytical formulations, we can obtain the corresponding
transmission coefficient which is also plotted in
Fig.~\ref{fig_dispersion1} for comparison. A good correspondence is
shown and at high-wave-vector region, the discrepancy is cause by
insufficient sampling points as for the case of using large cell
size (e.g. $\Delta x>\lambda/10$) for conventional FDTD.

\begin{figure}[t]
\centering
\includegraphics[width=10cm]{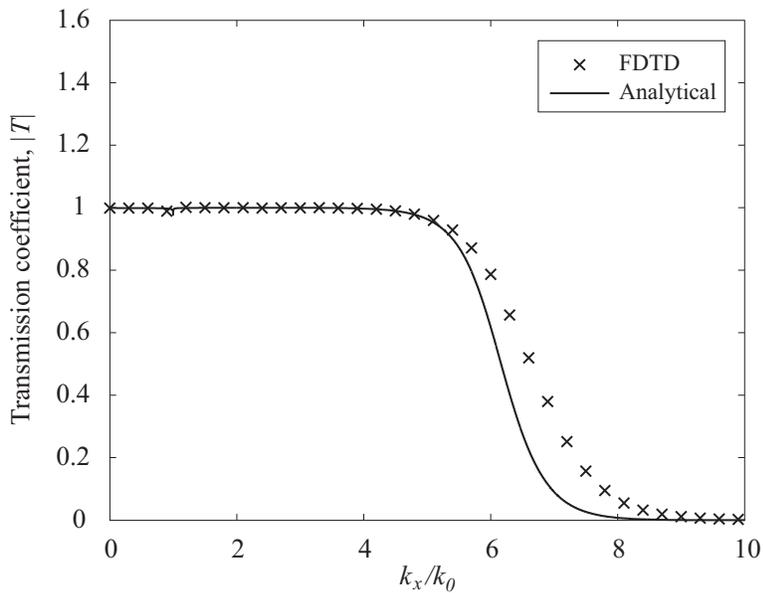}
\caption{Transmission coefficient of infinite planar LHM slabs using
the proposed FDTD method with averaged permittivity and with the
correction of material parameters. The numerical permittivity in
FDTD ($\Delta x=\lambda/40$) is $\varepsilon_r=-1-0.001j$. The same
permittivity is used to obtain the analytical solution for
comparison.} \label{fig_dispersion2}
\end{figure}

Another advantage of estimating the numerical permittivity is the
correction of the mismatch for FDTD simulations. After simple
derivations, we can obtain corrected plasma frequency and collision
frequency as
\begin{eqnarray}
\tilde{\omega}^2_p&=&\frac{2\sin\frac{\omega\Delta
t}{2}\left[-2(\varepsilon'_r-1)\sin\frac{\omega\Delta
t}{2}-\varepsilon''_r\gamma\Delta t\cos\frac{\omega\Delta
t}{2}\right]}{(\Delta t)^2\cos^2\frac{\omega\Delta
t}{2}},\nonumber\\
\tilde{\gamma}&=&\frac{2\varepsilon''_r\sin\frac{\omega\Delta
t}{2}}{(\varepsilon'_r-1)\Delta t\cos\frac{\omega\Delta t}{2}},
\label{eq_corrected}
\end{eqnarray}
where $\varepsilon'_r$ and $\varepsilon''_r$ are the real and
imaginary parts of the design relative permittivity $\varepsilon_r$,
respectively. For the case of $\varepsilon_r=-1-0.001j$, substitute
$\varepsilon'_r=-1$ and $\varepsilon''_r=-0.001$ into
(\ref{eq_corrected}) we get $\tilde{\omega}_p=1.4157\omega$ and
$\tilde{\gamma}=5.0051\times10^{-4}\omega$. Using the corrected
material parameters, the FDTD simulation result and its comparison
with analytical solutions are shown in Fig. \ref{fig_dispersion2}.
It can be seen that the mismatch has been canceled in FDTD
simulations hence there is no amplification in transmission
coefficient. Again the discrepancy with exact solutions in
high-wave-vector region is caused by insufficient sampling points
for such an FDTD spatial resolution of $\Delta x=\lambda/40$.
Therefore we suggest to use FDTD cell size smaller than $\Delta
x=\lambda/80$ for modelling of LHMs especially when evanescent waves
are involved.

\section{Effects of Switching Time}
Conventionally for single frequency simulations, the source should
be smoothly switched to its maximum value in order to avoid exciting
other frequency components \cite{ZiolkowskiPRE}. For modelling of
LHMs, the switching time has even more significant effect on their
behaviour. It is well known that the switching time considerably
influences the oscillation of images and often thirty period is used
as the switching time \cite{Cummer,Rao}. However, perhaps this is
the reason that no stable images could be obtained in
\cite{ZiolkowskiPRE,Rao} since recently, it is reported in
\cite{HuangSwitching} that using a switching time of at least one
hundred periods one can obtain stabilised image for lossless LHMs.

\begin{figure}[t]
\centering
\includegraphics[width=10cm]{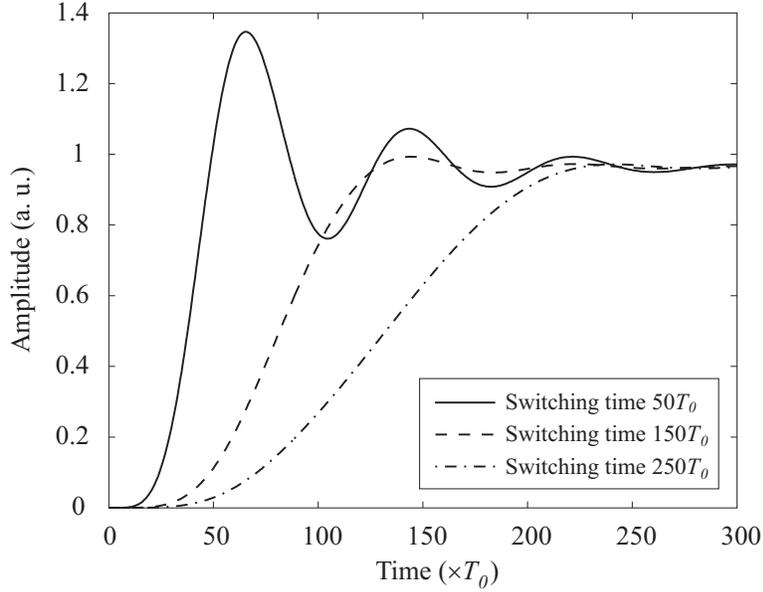}
\caption{The influence of different switching time on the
convergence time in FDTD simulation of infinite LHM slabs for a
fixed wave number $k_x=3k_0$. The $T_0$ is the period of the
sinusoidal wave function at the operating frequency. The field
intensity is taken at the second image plane of the LHM slab.}
\label{fig_switching}
\end{figure}

In our FDTD simulations, we also notice that switching time
influences the oscillation of the field intensity at the image plane
and hence the convergence time in simulations. We have performed
FDTD simulations with different switching time equal to $50T_0$,
$150T_0$ and $250T_0$ where $T_0$ is the period of the sinusoidal
signal. The FDTD cell size is $\Delta x=\lambda/100$ and corrected
material parameters from (\ref{eq_corrected}) are used. In order to
ensure faster convergence, we have chosen larger amount of losses
and used $\tilde{\varepsilon}_r=-1-0.01j$ in simulations. It should
be noted that because high-wave-vector components travel very slowly
in LHM slabs and the process of the growth of evanescent waves
requires a very long time to reach the steady-state, field values
should be taken only after total convergence is reached in
simulations. For our case of $\tilde{\varepsilon}_r=-1-0.01j$, we
have used a criteria of 0.001\% for detecting iteration errors and
terminating simulations. It is clearly shown in
Fig.~\ref{fig_switching} that for a fixed wave number ($k_x=3k_0$),
the oscillation of field intensity can be significantly suppressed
by prolonging the switching time.

\begin{figure}[t]
\centering
\includegraphics[width=10cm]{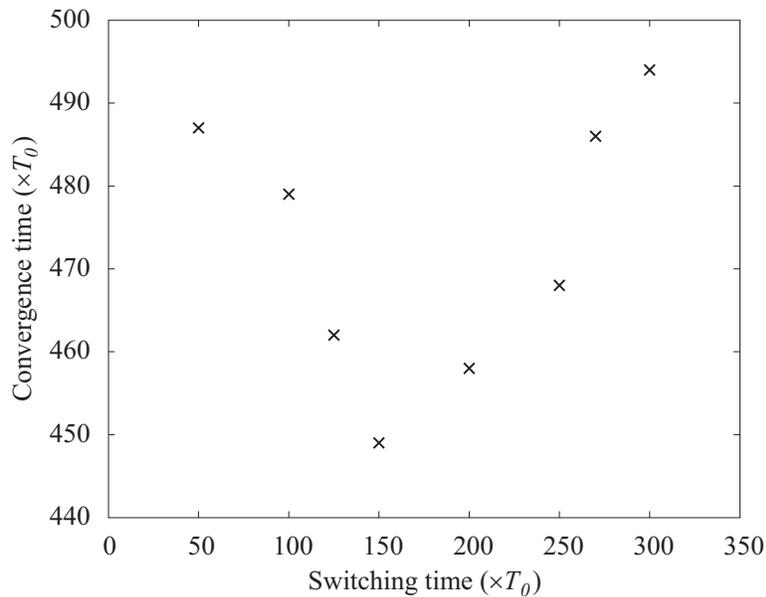}
\caption{The dependence of the convergence time on the switching
time in FDTD simulations of infinite LHM slabs for a fixed wave
number $k_x=3k_0$. A criteria of 0.001\% is used to detect iteration
errors and terminating simulations.} \label{fig_convergence}
\end{figure}

It is understandable that when the oscillation can be neglected, the
convergence time increases with the switching time. For
demonstration of the impact of switching time on convergence time,
we have performed FDTD simulations with various switching time. The
collected data is plotted in Fig.~\ref{fig_convergence}.
It can be seen that there exists an optimum switching time when the
minimum convergence time can be achieved for the case of $k_x=3k_0$.
However for different wave vectors and different material
parameters, the behaviour of oscillation differs considerably and in
certain cases the oscillation may last until a very long time. For
practical simulations such as modelling of subwavelength imaging by
a line source, since the source contains all wave vectors, therefore
it is necessary to switch the source slowly enough to ensure and
speed up the convergence of simulations.

\section{Conclusions}
In conclusion, we have performed simulations of LHMs using the
dispersive FDTD method. Two ADE methods namely the (\textbf{E},
\textbf{D}, \textbf{H}, \textbf{B}) scheme and the (\textbf{E},
\textbf{J}, \textbf{H}, \textbf{M}) scheme which lead to exactly
same results and the respective averaging techniques along material
boundaries are introduced. The comparison with exact analytical
solutions demonstrates that the averaging of
permittivity/permeability along the boundaries of LHM slabs is
essential for correct and accurate modelling of LHMs. The numerical
permittivity in FDTD is formulated where a mismatch between
numerical and analytical permittivity is introduced by FDTD time
discretisation. We suggest to correct such a mismatch in order to
model LHMs with their desired parameters in FDTD. The behaviour of
oscillation of field intensity for different switching time is also
analysed. It is shown that there exists an optimum value which leads
to fast convergence in simulations.

\end{document}